\documentclass[11pt]{article}

\usepackage[final]{acl}

\usepackage{times}
\usepackage{latexsym}

\usepackage[T1]{fontenc}

\usepackage[utf8]{inputenc}

\usepackage{microtype}
\usepackage{url}
\usepackage[breakable]{tcolorbox}
\usepackage{CJKutf8}
\newcommand{\zh}[1]{\begin{CJK*}{UTF8}{gbsn}\small #1\end{CJK*}}
\newcommand{\ja}[1]{\begin{CJK*}{UTF8}{min}\small #1\end{CJK*}}
\usepackage{inconsolata}
\usepackage{amsmath}
\usepackage{amsfonts}
\usepackage{enumitem}
\usepackage{arydshln}

\usepackage{graphicx}
\usepackage{hyperref}
\usepackage{multirow}
\usepackage{booktabs}
\usepackage{tabularx}
\usepackage{makecell}
\usepackage{siunitx}
\usepackage{subcaption}

\usepackage{listings}
\usepackage{xcolor}

\title{Code-Switching Information Retrieval: Benchmarks, Analysis, and the Limits of Current Retrievers}

\author{
  Qingcheng Zeng$^{1}$\thanks{Equal contribution.}, Yuheng Lu$^{2}$\footnotemark[1], Zeqi Zhou$^{3}$, Heli Qi$^{2,4}$, Puxuan Yu$^{5}$, \\\textbf{Fuheng Zhao}$^{6}$, \textbf{Hitomi Yanaka}$^{7,4}$, \textbf{Weihao Xuan}$^{7,4}$\thanks{Corresponding author.}, \textbf{Naoto Yokoya}$^{7,4}$ \\
  $^{1}$Northwestern University, $^{2}$Waseda University, $^{3}$Brown University \\ $^{4}$ RIKEN AIP, $^{5}$Snowflake Inc., $^{6}$University of Utah, $^{7}$The University of Tokyo\\
}

\begin{document}
\maketitle
\begin{abstract}
Code-switching is a pervasive linguistic phenomenon in global communication, yet modern information retrieval systems remain predominantly designed for, and evaluated within, monolingual contexts. To bridge this gap, we present a holistic study of code-switching IR. We introduce the \textbf{C}ode-\textbf{S}witching \textbf{R}etrieval benchmark-\textbf{L}ite (\textbf{CSR-L}), a human-annotated benchmark designed to capture natural mixed-language queries, and evaluate statistical, dense, cross-encoder, and late-interaction retrieval methods on it. The results show that code-switching is a persistent performance bottleneck, degrading even strong multilingual models. We further show that this failure is associated with substantial divergence between monolingual and code-switched query embeddings. To test whether the pattern generalizes beyond retrieval, we construct \textbf{CS-MTEB}, a benchmark covering 11 diverse tasks, where performance drops reach up to 27\%. Finally, we examine lexicon-based vocabulary expansion and find that, while it yields partial gains, it does not close the gap to monolingual performance. These findings underscore the fragility of current systems and establish code-switching as a crucial frontier for future IR optimization.
\end{abstract}

\section{Introduction}
Information retrieval (IR) stands as a cornerstone infrastructure for a wide array of intelligent applications, serving as the backbone for modern search engines, retrieval-augmented generation (RAG) systems \cite{lewis2021retrievalaugmentedgenerationknowledgeintensivenlp}, and autonomous search agents \cite{jin2025searchr1trainingllmsreason, zhao2025access}. Its ability to efficiently locate relevant data is critical for grounding generative models and enabling users to access vast repositories of knowledge. The underlying algorithms powering IR have undergone a significant evolution, shifting from traditional statistical methods like BM25 \cite{BM25} to semantic-aware dense retrieval \cite{gao-etal-2021-simcse} and sophisticated late interaction architectures \cite{ColBERT}. Crucially, as digital information becomes increasingly globalized, the field has expanded far beyond English-centric approaches. We have witnessed a vital shift toward robust multilingual IR \cite{yu2024arcticembed20multilingualretrieval} and complex cross-lingual IR \cite{zuo-etal-2025-evaluating}, which are essential for processing the diverse linguistic landscapes of the real world.

Despite the extensive evaluation of IR across multiple languages, one pervasive linguistic phenomenon remains critically understudied in current literature: code-switching \cite{code-switching}. This omission is striking given that code-switching is a fundamental aspect of global communication, particularly as approximately 70\% of the world population consists of bilingual speakers \cite{bilingual_reader}. Sociolinguistic studies highlight this frequency; for instance, \citet{Ahmed2024CodeSwitching} investigated speech communities in three countries and observed that code-switching occurs more than 15 times every 10 minutes. Within the context of search, \citet{mixed_script} conducted a large-scale analysis of Microsoft Bing logs and identified a substantial volume of code-switching queries. This trend was notably pronounced in the entertainment domain, where mixed-language inputs constituted around 27\% of overall traffic. Collectively, these findings underscore the urgent need to address code-switching in retrieval systems. Yet, we still lack a systematic evaluation of code-switching IR capabilities.

\begin{figure*}
    \centering
    \includegraphics[width=\textwidth]{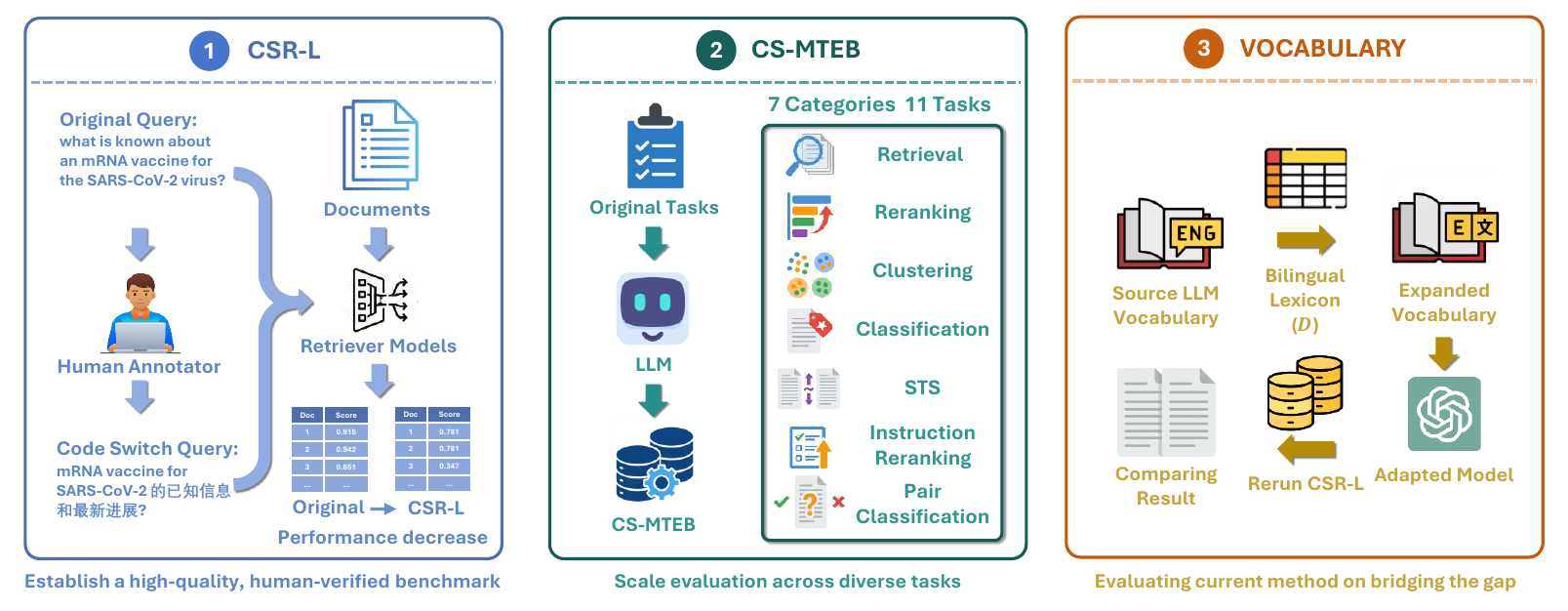}
    \caption{Overview of our comprehensive study on Code-Switching IR. Our framework proceeds in three stages: (1) CSR-L: We establish a high-quality, human-verified retrieval benchmark to assess natural mixed-language queries. (2) CS-MTEB: We scale the evaluation to 11 diverse tasks across 7 categories using LLM-assisted generation. (3) Vocabulary Expansion: We investigate lexicon-based vocabulary adaptation as a strategy to bridge the embedding space divergence between pure and code-switched text.}
    \label{fig:overall_teaser}
\end{figure*}

In this paper, we present the first holistic study of code-switching IR. Our framework, summarized in \autoref{fig:overall_teaser}, proceeds in three stages. First, we build the \textbf{C}ode-\textbf{S}witching \textbf{R}etrieval benchmark-\textbf{L}ite (\textbf{CSR-L}), a human-annotated benchmark that captures natural code-switched queries, and evaluate statistical, dense, cross-encoder, and late-interaction retrieval methods on it. This analysis shows that even simple query-side code-switching substantially degrades retrieval quality, including for strong multilingual retrievers, and that the degradation is accompanied by a clear shift in embedding space. Second, we scale the study beyond standard retrieval by introducing \textbf{CS-MTEB}, an MTEB-style benchmark covering 11 tasks across 7 task types; across these tasks, advanced embedding models still exhibit performance drops of up to 27\%. Third, we test whether lexicon-based vocabulary expansion can mitigate the problem. Although this intervention improves English-centric retrievers, it still falls short of restoring monolingual performance. Taken together, these results identify code-switching as a major robustness gap in current IR systems. Code and datasets are publicly available in our \href{https://github.com/paddler2022/Code-Switching-Information-Retrieval}{GitHub repository} and the \href{https://huggingface.co/collections/UTokyo-Yokoya-Lab/cs-mteb}{CS-MTEB} and \href{https://huggingface.co/collections/UTokyo-Yokoya-Lab/csr-l}{CSR-L} Hugging Face collections.

\section{Related Work}
\paragraph{IR and Embedding Models Evaluation} The field of IR has undergone a fundamental transformation in its backbone methodology, evolving from lexical matching to semantic representation. In modern applications, the mainstream IR pipeline typically adopts a "retrieve-then-rerank" architecture to balance efficiency and precision. For the initial retrieval step, the paradigm has shifted toward dense retrieval, where most embedding models are trained using contrastive learning in a bi-encoder fashion. This approach encodes queries and documents into independent vector spaces, allowing for efficient similarity calculation via dot product or cosine similarity during inference. Training these models often involves sophisticated negative sampling strategies and loss functions, such as InfoNCE \cite{oord2018representation}, to optimize the separation between relevant and irrelevant passages. Following retrieval, a reranking stage is often employed—frequently utilizing cross-encoders—to re-score the top candidates with finer granularity by capturing the full interaction between query and document tokens. To rigorously assess these advancements, the community has developed a wide range of benchmarks. Early efforts like BEIR \cite{thakur2021beirheterogenousbenchmarkzeroshot} focused on measuring zero-shot generalization across diverse domains, while MTEB \cite{muennighoff2023mtebmassivetextembedding} expanded the scope to massive text embedding tasks beyond just retrieval. More recently, benchmarks such as BRIGHT \cite{su2025brightrealisticchallengingbenchmark} have been proposed to test models on highly challenging, realistic queries that require deep reasoning, pushing the boundaries of current embedding capabilities.

\paragraph{Multilingual and Cross-lingual Retrieval} Multilingual and cross-lingual retrieval performance of embedding models has received increasing attention. For example, MMTEB \cite{enevoldsen2025mmtebmassivemultilingualtext} evaluated embedding models in over 250 languages and across more than 500 tasks. \citet{litschko-etal-2025-cross} evaluated IR models on cross-dialect retrieval. For training, \citet{wang2024multilingual,yu2024arcticembed20multilingualretrieval,zhang2025qwen3embeddingadvancingtext} represent some recent attempts to build multilingual retrievers using open-source and synthetic data. However, one crucial linguistic phenomenon, code-switching, remains relatively underexplored. \citet{litschko-etal-2023-boosting, do-etal-2024-contrastivemix} represent two preliminary attempts to use code-switching data to enhance multilingual and cross-lingual IR. Although \citet{winata-etal-2024-miners, kim-etal-2025-milq} touches on code-switching evaluation, it remains task- and setting-specific (e.g., sentiment analysis and bitext retrieval, focusing on late-interaction models) and does not provide a holistic picture of code-switching in embedding-based IR, which we address in this paper.

\section{Code-Switching Retrieval Benchmark-Lite (CSR-L)}
The naturalness of code-switched text has been examined from both theoretical \cite{poplack2020sometimes, myers1997duelling} and empirical \cite{pratapa-etal-2018-language, hsu-etal-2023-code} perspectives. However, the field remains without a single standard or automatic metric to reliably judge the naturalness of code-switching, which severely limits the scalability of benchmarks. Consequently, in this section, we employ human annotators to rewrite queries within IR benchmarks. This approach allows us to overcome the limitations of automated metrics, ensuring high data quality and facilitating a more reliable evaluation.

\subsection{Building CSR-L}
We selected four representative datasets containing a limited number of queries to facilitate rewriting: (1) Touché 2020 \cite{touche} for argument retrieval; (2) HumanEval \cite{chen2021codex} for code retrieval; (3) TRECCOVID \cite{roberts2021searchingscientificevidencepandemic} for biomedical IR; and (4) FollowIR \cite{weller-etal-2025-followir} for evaluating instruction-following capabilities. As these datasets are originally English-only, we rewrote the queries to introduce code-switching in two languages: Mandarin Chinese and Japanese.

Three authors of this paper participated in the query rewriting task. All are native Chinese speakers with professional proficiency in both English and Japanese, developed through their undergraduate and postgraduate education. The rewriting process followed two steps: (1) one annotator first rewrote the query into a code-switched form; and (2) a second annotator validated the result, with the authority to edit the text or discard the rewrite when necessary. The detailed instructions are provided in \autoref{sec:instruction_to_annotator}. Statistics for the final Chinese dataset are presented in \autoref{tab:statistics_of_Chinese_CSRL}, while the Japanese statistics are reported in \autoref{tab:statistics_of_Japanese_CSRL} in the appendix.

\begin{table}[!t]
\centering
\resizebox{\columnwidth}{!}{
\begin{tabular}{l|rrr|rr|c}
\toprule
& \multicolumn{3}{c|}{\textbf{Total Number}}
& \multicolumn{2}{c|}{\textbf{Avg. Length}}
& \textbf{Examples} \\
\cmidrule{2-4}\cmidrule{5-6}\cmidrule{7-7}
\textbf{Dataset}
& $\mathbf{Q}$
& $\boldsymbol{\mathcal{D}}$
& $\boldsymbol{\mathcal{D}^{+}}$
& $\mathbf{Q}$
& $\boldsymbol{\mathcal{D}}$
&  \\
\midrule
Touch\'e 2020     & 49 & 303,732 & 34.94 & 16.82 & 451.51 & \autoref{tab:touche2020_cs_example_cn} \\
HumanEval       & 158 & 158 & 1.00 & 64.76 & 98.20 & \autoref{tab:humanEval_cs_example_cn} \\
TRECCOVID   & 50 & 171,332 & 493.46 & 24.36 & 223.51 & \autoref{tab:treccovid_cs_example_cn} \\
FollowIR  & 198 & 98,312 & 30.07 & 111.15 & 465.39 & \autoref{tab:followIR_cs_example_cn} \\
\bottomrule
\end{tabular}
}
\caption{Statistics of datasets in CSR-L-Chinese. \textit{Q}: number of queries; \textit{D}: corpus size; \textit{D}$^+$: average positive documents per query. Avg. Length is measured in the GPT-2 \cite{radford2019language} tokenizer. Our query examples can be seen in the tables in Appendix.}
\label{tab:statistics_of_Chinese_CSRL}
\end{table}

\subsection{Evaluation Setup}
We evaluate CSR-L with four families of IR methods: (1) the lexical baseline BM25 \cite{BM25}; (2) bi-encoder retrievers, including \textit{all-MiniLM-L12-v2} \cite{reimers-gurevych-2019-sentence}, \textit{e5-large-v2} \cite{wang2024textembeddingsweaklysupervisedcontrastive} and \textit{multilingual-e5-large (mE5-large)} \cite{wang2024multilingual}, \textit{bge-m3} \cite{chen2024bge}, \textit{Arctic-Embed-m/l-v2.0} \cite{yu2024arcticembed20multilingualretrieval}, and \textit{Qwen3-Embedding-0.6/4/8B} \cite{zhang2025qwen3embeddingadvancingtext}; (3) cross-encoder rerankers, including \textit{jina-reranker-v3} \cite{wang2025jinarerankerv3lateinteractionlistwise}, \textit{bge-reranker-v2-m3} \cite{chen2024bge}, and \textit{Qwen3-Reranker-0.6/4/8B} \cite{zhang2025qwen3embeddingadvancingtext}; and (4) the late-interaction retriever \textit{ColBERT v2} \cite{santhanam2022colbertv2effectiveefficientretrieval}.

We use nDCG@10 as the primary metric throughout the evaluation, with the exception of FollowIR, where we report pairwise-MRR (\textit{p-MRR}). For each method, we compare performance on the original queries and their code-switched counterparts. For the cross-encoder results in CSR-L, we score each query--document pair directly over the full document set, rather than reranking a top-$k$ candidate pool produced by a separate first-stage retriever. Accordingly, the absolute cross-encoder numbers in \autoref{tab:csrl_results_ndcg10_avg_chinese} and \autoref{tab:csrl_results_ndcg10_avg_japanese} should be interpreted as direct full-corpus scoring results rather than as conventional two-stage reranking performance. Additional metric details are provided in \autoref{sec:evaluation_details}.

\section{CSR-L Results}

\begin{table*}[!t]
\centering
\small
\setlength{\tabcolsep}{5pt}

\resizebox{\textwidth}{!}{%
\begin{tabular}{l l *{8}{S[table-format=2.2]} *{2}{S[table-format=2.2]} S[table-format=-2.2]}
\toprule
\multirow{2}{*}{\textbf{Method Family}} &
\multirow{2}{*}{\textbf{Model}} &
\multicolumn{2}{c}{\textbf{Touch\'e 2020}} &
\multicolumn{2}{c}{\textbf{HumanEval}} &
\multicolumn{2}{c}{\textbf{TRECCOVID}} &
\multicolumn{2}{c}{\textbf{FollowIR}} &
\multicolumn{1}{c}{\textbf{Avg}} &
\multicolumn{1}{c}{\textbf{Avg}} &
\multicolumn{1}{c}{\textbf{Drop}} \\
\cmidrule(lr){3-4}\cmidrule(lr){5-6}\cmidrule(lr){7-8}\cmidrule(lr){9-10}\cmidrule(lr){11-13}
& &
{\textbf{Orig}} & {\textbf{CSR-L}} &
{\textbf{Orig}} & {\textbf{CSR-L}} &
{\textbf{Orig}} & {\textbf{CSR-L}} &
{\textbf{Orig}} & {\textbf{CSR-L}} &
{\textbf{Orig}} & {\textbf{CSR-L}} &
{$\Delta$} \\
\midrule

\multirow{1}{*}{Statistical} &
BM25 &
{60.32} & {37.68} & {35.02} & {41.79} & {55.62} & {46.43} & {-0.62} & {-1.8} & {37.59} & {31.03} & {-6.56} \\
\midrule

\multirow{9}{*}{Bi-encoder} &
\textit{e5-large-v2} &
{42.52} & {22.88} & {80.70} & {72.93} & {66.64} & {50.42} & {-0.99} & {-4.97} & {47.22} & {35.32} & {-11.90} \\
& \textit{all-MiniLM-L12-v2} &
{49.22} & {23.85} & {70.08} & {60.37} & {51.17} & {39.51} & {-0.66} & {-3.36} & {42.45} & {30.09} & {-12.36} \\
& \textit{mE5-large} &
{49.32} & {42.75} & {81.15} & {74.04} & {71.56} & {56.54} & {-3.38} & {-2.28} & {49.66} & {42.76} & {-6.90} \\
& \textit{bge-m3} &
{55.02} & {50.00} & {61.33} & {59.26} & {54.70} & {52.32} & {-2.94} & {-3.00} & {42.03} & {39.65} & {-2.38} \\
& \textit{Arctic-Embed-m-v2.0} &
{65.29} & {48.46} & {78.7} & {75.05} & {80.45} & {74.15} & {-3.20} & {-4.32} & {55.31} & {48.34} & {-6.97} \\
& \textit{Arctic-Embed-l-v2.0} &
{64.05} & {54.91} & {71.27} & {68.94} & {83.63} & {76.99} & {-2.45} & {-2.47} & {54.13} & {49.59} & {-4.54} \\
& \textit{Qwen3-Embedding-0.6B} &
{71.65} & {61.30} & {94.24} & {94.43} & {89.43} & {81.66} & {5.10} & {4.07} & {65.11} & {60.37} & {-4.74} \\
& \textit{Qwen3-Embedding-4B} &
{75.07} & {66.67} & {98.12} & {96.17} & {92.95} & {88.67} & {11.87} & {8.91} & {69.50} & {65.11} & {-4.39} \\
& \textit{Qwen3-Embedding-8B} &
{75.77} & {68.55} & {99.22} & {98.90} & {94.68} & {89.72} & {9.86} & {7.63} & {69.88} & {66.20} & {-3.68} \\
\midrule

\multirow{5}{*}{Cross-encoder} &
\textit{jina-reranker-v3} &
{22.68} & {24.96} & {85.53} & {84.43} & {81.32} & {68.07} & {-0.27} & {-0.17} & {47.32} & {44.32} & {-3.00} \\
& \textit{bge-reranker-v2-m3} &
{35.48} & {27.86} & {43.74} & {49.77} & {79.00} & {67.17} & {-1.38} & {0.32} & {39.21} & {36.28} & {-2.93} \\
& \textit{Qwen3-Reranker-0.6B} &
{29.15} & {23.91} & {83.74} & {84.11} & {84.30} & {71.19} & {1.40} & {-0.01} & {49.65} & {44.80} & {-4.85} \\
& \textit{Qwen3-Reranker-4B} & {37.76} & {28.34} & {85.29} & {84.11} & {85.44} & {70.86} & {2.33} & {-1.01} & {52.71} & {45.58} & {-7.13} \\
& \textit{Qwen3-Reranker-8B} & {40.91} & {32.01} & {85.53} & {84.62} & {84.58} & {69.88} & {2.74} & {0.56} & {53.44} & {46.77} & {-6.67} \\
\midrule

\multirow{1}{*}{Late-interaction} &
ColBERT v2 &
{61.62} & {29.30} & {40.30} & {42.46} & {69.30} & {53.74} & {-0.95} & {-0.46} & {42.57} & {31.26} & {-11.31} \\

\bottomrule
\end{tabular}%
}

\vspace{2pt}
\caption{nDCG@10 and \textit{p-MRR} on the original (Orig) and code-switched (CSR-L) queries across four IR benchmarks on English-Chinese code-switching. Avg is the macro-average over the four datasets. Drop $\Delta$ is computed as Avg(CSR-L) - Avg(Orig); negative values indicate performance degradation under code-switching.}
\label{tab:csrl_results_ndcg10_avg_chinese}
\end{table*}

\subsection{General Results}
The Chinese results are shown in \autoref{tab:csrl_results_ndcg10_avg_chinese}, while the Japanese results are reported in \autoref{tab:csrl_results_ndcg10_avg_japanese} in \autoref{sec:csrl_results_on_japanese}. The overall pattern is highly consistent across the two languages. First, query-side code-switching alone substantially degrades performance on the main retrieval datasets, even though the underlying documents remain unchanged. The newly added multilingual bi-encoder baselines, \textit{mE5-large} and \textit{bge-m3}, follow the same trend, showing that multilingual encoders are more robust but not immune.
The degradation is especially large on Touch\'e 2020 and TRECCOVID, whereas it is milder on HumanEval, likely because that benchmark is structurally simpler. Among English-centric bi-encoders such as \textit{e5-large-v2}, the drop reaches roughly 15 points on the two general retrieval datasets. Even for the \textit{Qwen3-Embedding} series, which is comparatively more robust, the decrease on Touch\'e 2020 and TRECCOVID still exceeds 8 points in some settings. Model scaling helps, but even the 8B variant does not eliminate the gap.

\begin{figure*}[!t]
    \centering
    \resizebox{\textwidth}{!}{%
        \begin{subfigure}[t]{0.24\textwidth}
            \centering
            \includegraphics[width=\linewidth]{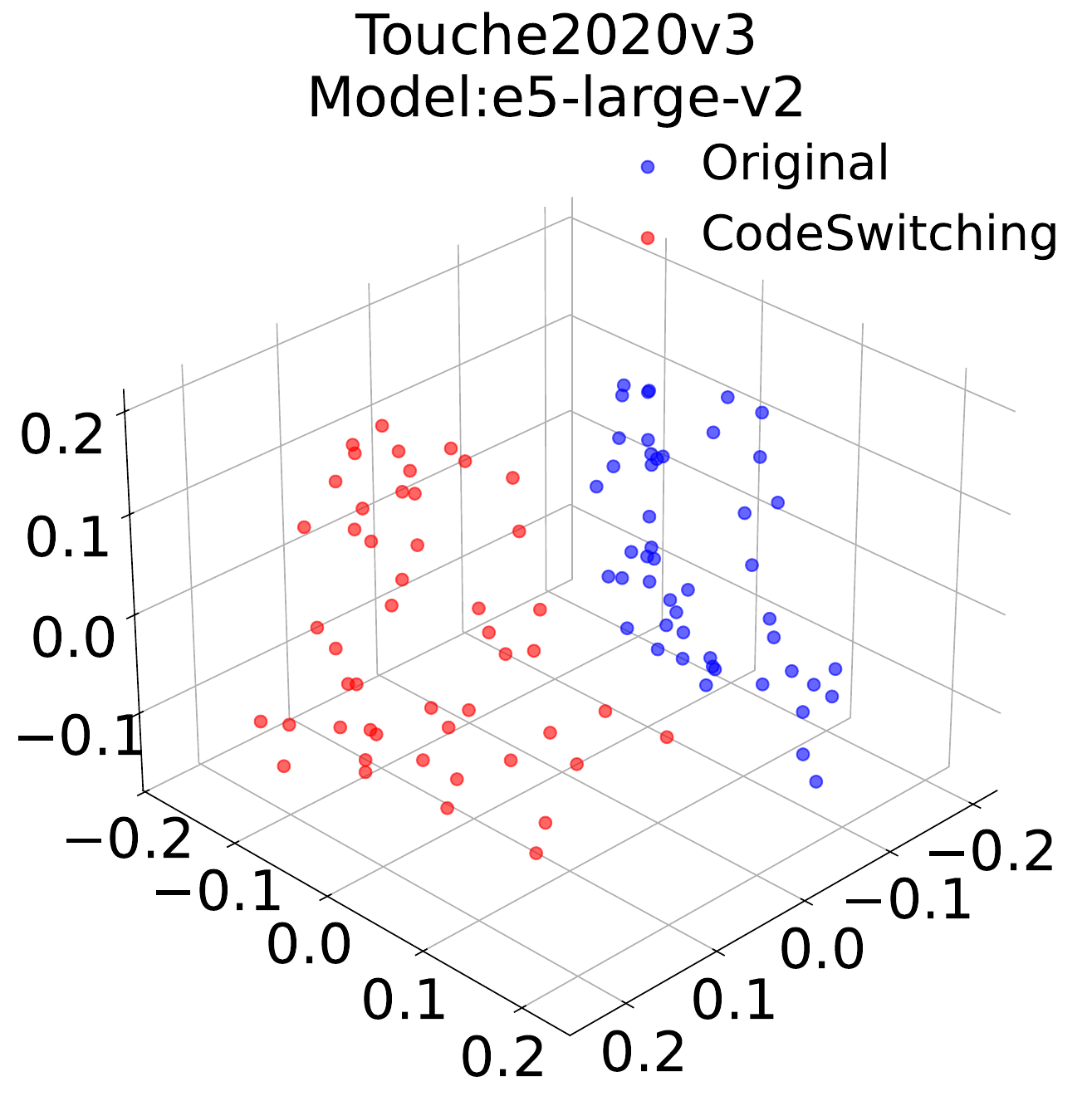}
            \caption{\textit{e5} on Touch\'e}
            \label{fig:sub-a}
        \end{subfigure}
        \hfill
        \begin{subfigure}[t]{0.24\textwidth}
            \centering
            \includegraphics[width=\linewidth]{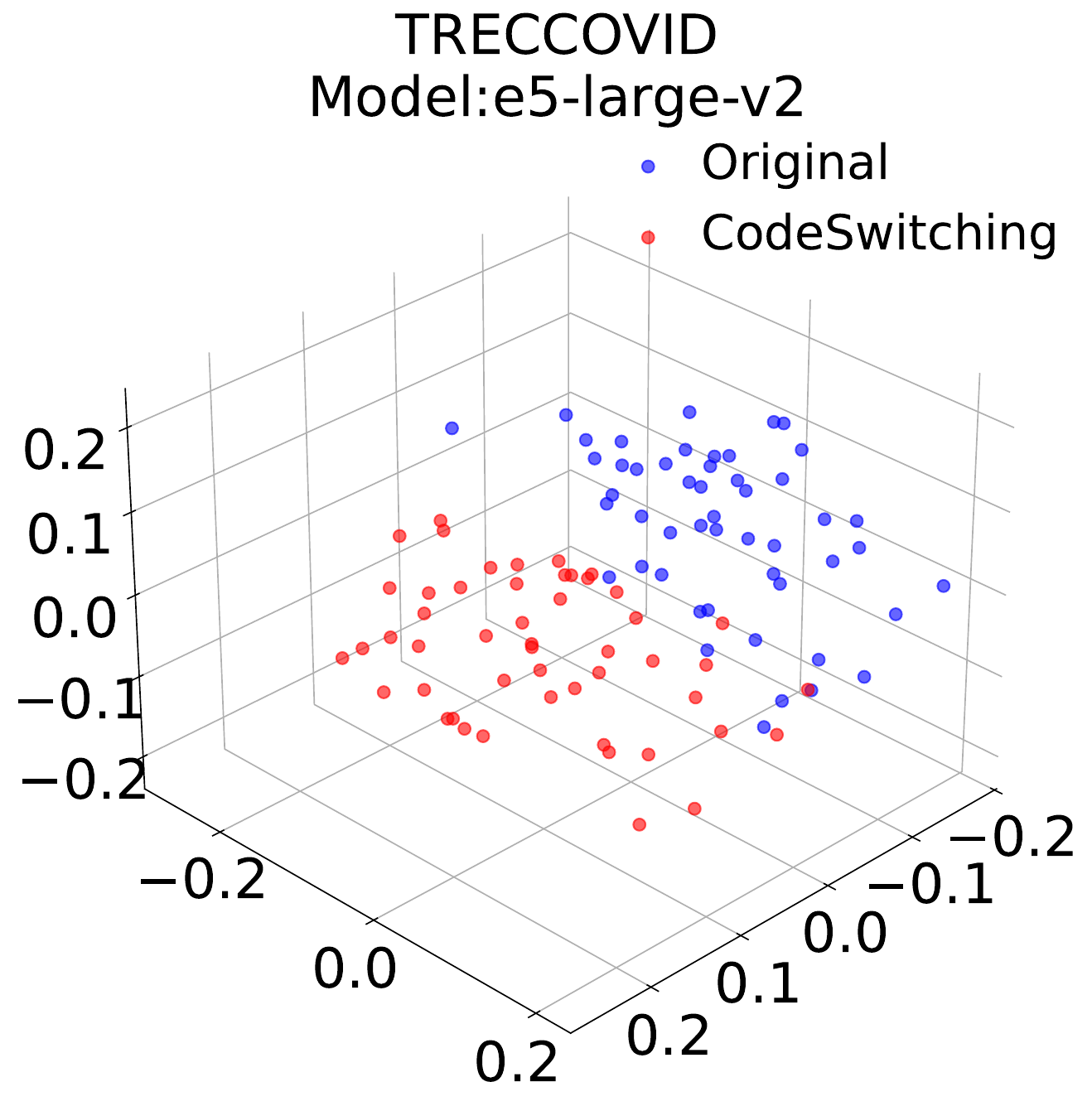}
            \caption{\textit{e5} on TREC}
            \label{fig:sub-b}
        \end{subfigure}
        \hfill
        \begin{subfigure}[t]{0.24\textwidth}
            \centering
            \includegraphics[width=\linewidth]{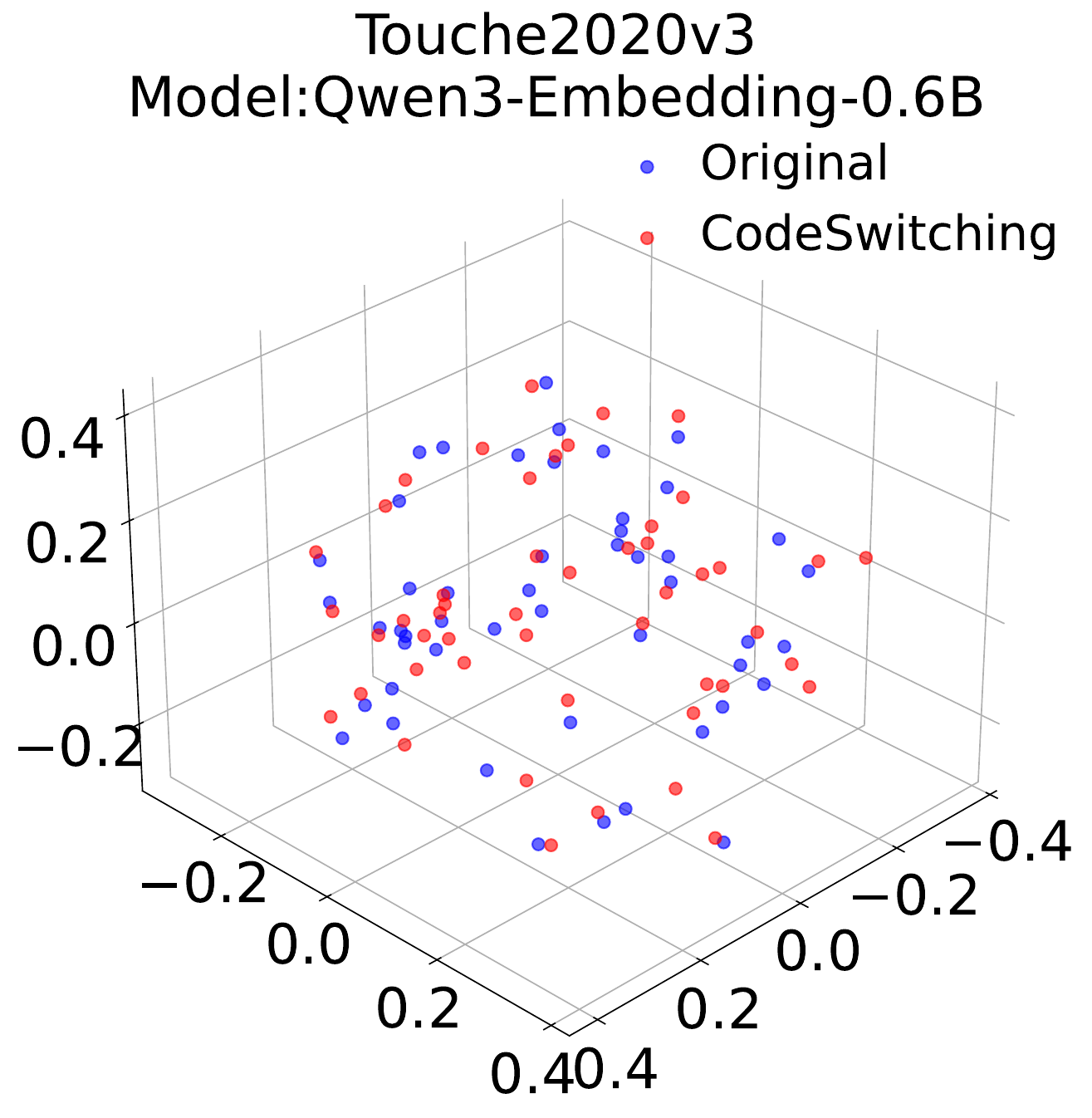}
            \caption{\textit{Qwen 0.6B} on Touch\'e}
            \label{fig:sub-c}
        \end{subfigure}
        \hfill
        \begin{subfigure}[t]{0.24\textwidth}
            \centering
            \includegraphics[width=\linewidth]{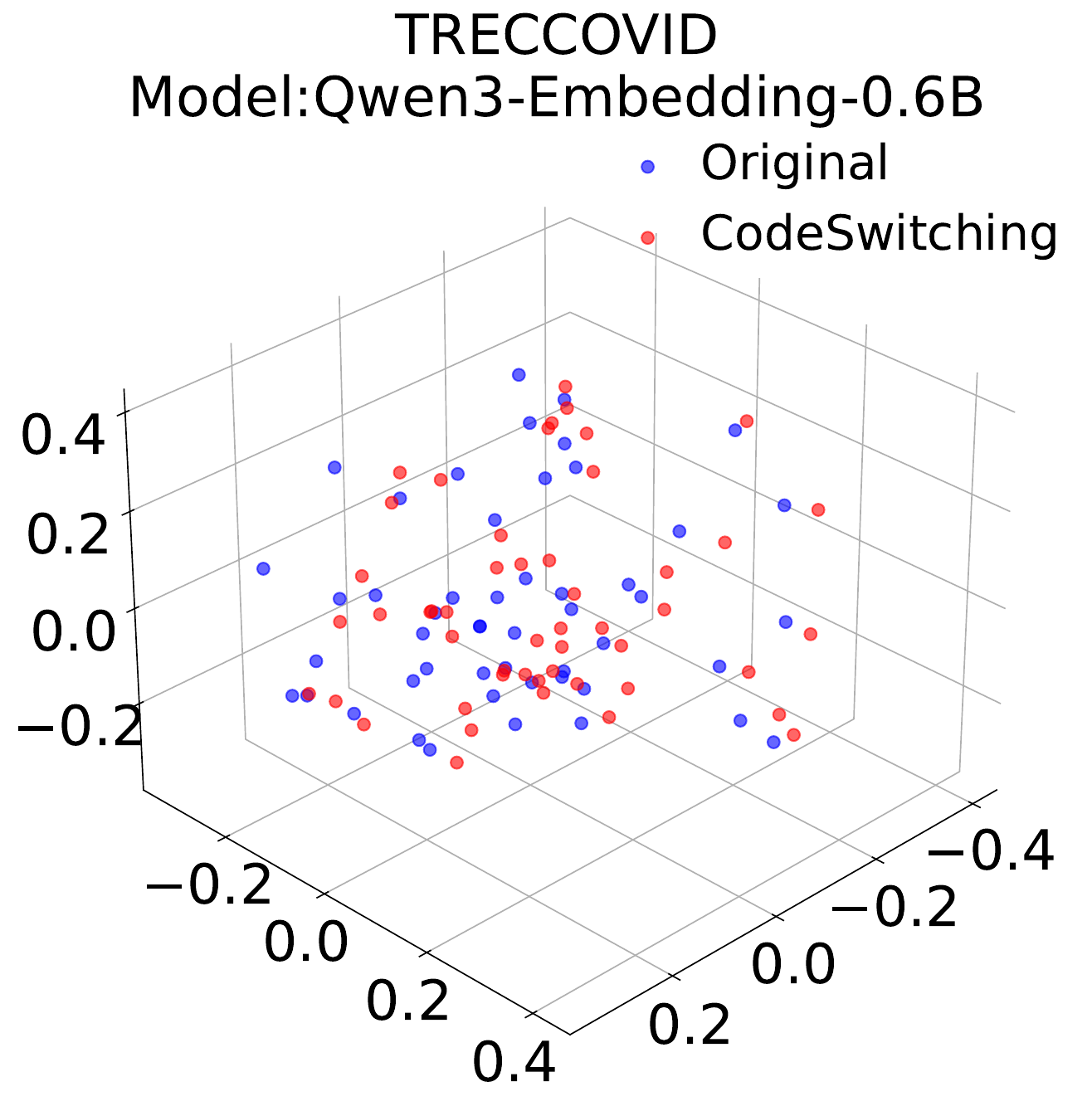}
            \caption{\textit{Qwen 0.6B} on TREC}
            \label{fig:sub-d}
        \end{subfigure}
    } 
    \caption{The visualization of \textit{e5} and \textit{Qwen 0.6B} embeddings on two IR datasets.}
    \label{fig:four-subfig}
\end{figure*}

While all evaluated models struggle, an important distinction emerges between English-centric systems and multilingual retrievers. Multilingual models generally exhibit a smaller relative decline than their English-only counterparts. For instance, when controlling for model size, \textit{Arctic-Embed-m-v2.0} experiences a substantially smaller drop than \textit{e5-large-v2}. This relative stability suggests that exposure to diverse languages during training provides a meaningful benefit, helping the model interpret code-switching patterns and partially absorb the disruption caused by linguistic mixing.

Finally, we observe no significant variation in robustness across different retrieval paradigms. For example, despite the higher computational cost associated with cross-encoders, these models do not exhibit superior resistance to the performance drops caused by code-switched queries. We note, however, that the cross-encoder scores in \autoref{tab:csrl_results_ndcg10_avg_chinese} and \autoref{tab:csrl_results_ndcg10_avg_japanese} come from direct full-corpus scoring rather than a standard retrieve-then-rerank pipeline, so their absolute values should not be read as directly comparable to conventional reranking benchmarks. This vulnerability is equally prevalent in statistical methods and late-interaction frameworks. Taken together, our results on CSR-L suggest that while multilingual pre-training offers partial mitigation, code-switching poses a fundamental challenge that neither architectural complexity nor current scaling strategies can fully overcome.

\subsection{Embedding Space Analysis}

Visualizing embedding spaces provides valuable insights into the underlying causes of retrieval failure. In this subsection, we focus on Touch\'e 2020 and TRECCOVID, two datasets where models exhibited significant performance degradation. We selected \textit{e5-large-v2} and \textit{Qwen3-Embedding-0.6B} as representative models and visualized their query representations in a three-dimensional space using Principal Component Analysis (PCA), as shown in \autoref{fig:four-subfig}.

Our analysis uncovers distinct geometric behaviors. For the English-centric retriever \textit{e5-large-v2}, code-switching induces a drastic shift in the embedding space: the original and code-switched queries separate into two disjoint dense clusters rather than forming a shared semantic distribution. In contrast, multilingual models exhibit greater stability. For example, on Touch\'e 2020, the centroid distance is smaller for \textit{Qwen3-Embedding-0.6B} than for \textit{e5-large-v2} (0.20 vs.\ 0.25), and the two query sets overlap much more strongly. This geometric resilience likely contributes to the more moderate performance declines observed in multilingual models. Nevertheless, the gap does not disappear, suggesting that code-switching introduces semantic difficulties that go beyond what standard multilingual pre-training currently resolves.

\begin{table*}[t]
\centering
\small
\setlength{\tabcolsep}{4pt}
\resizebox{\textwidth}{!}{
\begin{tabular}{llcccccccc}
\toprule
\textbf{Model} & \textbf{Setting} &
\textbf{Instr. Rerank} (1) &
\textbf{Retrieval} (5) &
\textbf{Clust.} (1) &
\textbf{Cls.} (1) &
\textbf{STS} (1) &
\textbf{Rerank} (1) &
\textbf{Pair Cls.} (1) &
\textbf{Total} (11) \\
\midrule

\multirow{5}{*}{e5-large-v2}
& Original & -0.99 & 51.78 & 62.00 & 73.97 & 84.55 & 60.17 & 59.88 & 55.91 \\
\cdashline{2-10}
& Chinese    & -4.97 & 40.49 & 55.71 & 57.47 & 59.42 & 22.39 & 54.42 & 40.70 \\
& Japanese    & -1.93 & 40.14 & 54.87 & 62.76 & 65.71 & 25.75 & 55.17 & 43.21 \\
& German    & -2.54 & 40.80 & 55.65 & 61.74 & 67.54 & 26.20 & 56.11 & 43.64 \\
& Spanish    & -2.31 & 41.49 & 59.03 & 63.06 & 69.55 & 26.38 & 57.79 & 45.00 \\
\midrule

\multirow{5}{*}{Arctic-Embed-m-v2.0}
& Original & -3.20 & 64.21 & 60.09 & 64.78 & 75.97 & 62.37 & 58.09 & 54.62 \\
\cdashline{2-10}
& Chinese    & -4.32 & 57.54 & 54.09 & 56.94 & 62.48 & 34.30 & 56.29 & 45.33 \\
& Japanese    & -3.13 & 57.79 & 54.45 & 58.80 & 64.73 & 37.15 & 56.43 & 46.60 \\
& German    & -3.83 & 60.39 & 53.05 & 62.66 & 68.72 & 37.09 & 57.62 & 47.98 \\
& Spanish    & -3.43 & 61.60 & 58.52 & 62.09 & 68.12 & 38.08 & 58.17 & 49.02 \\
\midrule

\multirow{5}{*}{Qwen3-Embedding-0.6B}
& Original & 5.10 & 73.67 & 68.21 & 72.07 & 91.14 & 63.09 & 75.55 & 64.12 \\
\cdashline{2-10}
& Chinese    & 4.07 & 69.69 & 61.70 & 68.50 & 86.69 & 37.13 & 72.90 & 57.24 \\
& Japanese    & 4.11 & 68.68 & 63.10 & 68.58 & 85.23 & 37.33 & 72.29 & 57.05 \\
& German    & 3.54 & 66.26 & 64.83 & 68.00 & 85.86 & 36.04 & 73.08 & 56.80 \\
& Spanish    & 3.75 & 69.80 & 63.77 & 68.71 & 86.16 & 36.20 & 74.25 & 57.52 \\
\bottomrule
\end{tabular}
}
\caption{CS-MTEB results by model and evaluation setting. Columns correspond to CS-MTEB task categories, with the number of tasks per category in parentheses. The result is the macro average over 7 task categories / Mean (TaskType).}
\label{tab:cs-mteb}
\end{table*}

\section{CS-MTEB}
In the preceding section, our results demonstrated that regardless of model scale, IR paradigm, or multilingual pre-training, current methods consistently fail to maintain parity with monolingual performance when processing code-switched queries. Recognizing the critical need to assess the universality of this deficit, we now expand our evaluation beyond standard retrieval tasks. In this section, we leverage LLMs to scale our investigation, covering a broader spectrum of task types, datasets, and language pairs. By aligning with the rigorous standards of the general-purpose MTEB benchmark, we introduce \textbf{CS-MTEB}. This comprehensive framework is designed to provide a holistic diagnosis of text embedding models, systematically uncovering the boundaries of their success and failure in mixed-language scenarios.

\subsection{Covered Tasks}
To ensure a comprehensive evaluation, we curated a diverse set of tasks for this benchmark, spanning the following categories:
\begin{itemize}[itemsep=0pt, topsep=0pt, parsep=0pt, partopsep=0pt]
    \item Instruction Reranking: We incorporate \textit{FollowIR} \cite{weller-etal-2025-followir}, aligning our setup with standard MTEB protocols.
    \item Retrieval: Beyond the three datasets established in CSR-L, we expand the scope by including \textit{Arguana} \cite{wachsmuth-etal-2018-retrieval} and \textit{ClimateFEVERHardNegatives} \cite{diggelmann2021climatefeverdatasetverificationrealworld}.
    \item Clustering: We utilize \textit{ArXivHierarchicalClusteringP2P} \cite{enevoldsen2025mmtebmassivemultilingualtext}. For this task, we randomly introduce code-switching into 10\% of the document text to quantify the resulting impact on clustering stability.
    \item Classification: We adopt the test set of \textit{TweetSentimentExtractionClassification} \cite{enevoldsen2025mmtebmassivemultilingualtext} to serve as the representative classification benchmark.
    \item Semantic Textual Similarity (STS): We employ the \textit{STS Benchmark} \cite{enevoldsen2025mmtebmassivemultilingualtext} to assess the models' semantic understanding. Specifically, we apply code-switching to one sentence within each pair to test cross-lingual alignment.
    \item Reranking: We leverage \textit{AskUbuntuDupQuestions} \cite{lei-etal-2016-semi} to evaluate reranking capabilities. Consistent with our retrieval setup, we apply code-switching exclusively to the query side.
    \item Pair Classification: We utilize \textit{TwitterSemEval2015} \cite{xu-etal-2015-semeval} for pair classification. Similarly, we introduce code-switching into one sentence of the pair to challenge the model's judgment.
\end{itemize}

By encompassing 11 distinct tasks across these 7 categories, we aim to construct a holistic picture of how code-switching influences the performance of text embedding models.

\subsection{Experimental Setup}
Because fully human rewriting is infeasible at MTEB scale, we use an LLM to generate the code-switched variants. We refined the prompt templates through several iterations, grounding the design in the human-authored CSR-L queries to preserve both naturalness and information need. The final prompt template is provided in \autoref{appendix:prompt_template_doing_code_switching}, and a manual quality check on 50 sampled rewritten queries is reported in \autoref{tab:csmteb_query_quality_check}.

We selected \texttt{MiMo-V2-Flash} \cite{coreteam2026mimov2flashtechnicalreport} as the backbone model for this generation task. Our evaluation incorporates 9 languages mixed with English, including Chinese, Japanese, German, Spanish, Korean, French, Italian, Portuguese, and Dutch. For the experimental analysis, we assess the following models: \textit{e5-large-v2} \cite{wang2024textembeddingsweaklysupervisedcontrastive}, \textit{Arctic-Embed-m-v2.0} \cite{yu2024arcticembed20multilingualretrieval}, and \textit{Qwen3-Embedding-0.6B} \cite{zhang2025qwen3embeddingadvancingtext}.

\subsection{Results}
The main results of CS-MTEB on four languages are presented in Table~\ref{tab:cs-mteb}, with an additional five languages reported in \autoref{tab:cs-mteb-additional-five-languages}. The same qualitative pattern from CSR-L reappears: across models, tasks, and language pairs, code-switching remains a broad and persistent bottleneck.

First, for English-centric models such as \textit{e5-large-v2}~\cite{wang2024textembeddingsweaklysupervisedcontrastive}, the performance degradation is substantial and consistent. We observe a significant drop in the average score across all four language mixtures, ranging from approximately 10 to 15 points. Critically, this decline occurs regardless of linguistic proximity; the model suffers similar losses whether English is mixed with typologically distinct languages like Chinese and Japanese, or with closer European relatives such as German and Spanish. This universality suggests that without explicit multilingual alignment, the embedding space is highly fragile to the semantic noise introduced by code-switching.

In contrast, the multilingual retriever \textit{Arctic-Embed-m-v2.0}~\cite{yu2024arcticembed20multilingualretrieval} exhibits greater resilience, although it is not immune. Across the same set of languages, the performance decline is noticeably mitigated compared to the monolingual baseline. For instance, in the Spanish-English setting, the model experiences a drop of approximately 5 points, compared to the $\sim$10 point drop observed in \textit{e5-large-v2}. While this indicates that exposure to diverse linguistic data provides a foundational robustness, the persistence of these gaps underscores that standard multilingual training alone is insufficient to fully bridge the code-switching deficit.

Analyzing performance across different task categories reveals distinct sensitivities. While retrieval tasks exhibit a consistent decline, finer-grained objectives can be substantially more fragile, with reranking showing the sharpest failures. For example, in the Japanese code-switching setting, \textit{e5-large-v2} suffers a catastrophic degradation in reranking, plummeting from a baseline of 60.17 to just 25.75. In contrast, more decision-oriented tasks such as pair classification tend to be comparatively less sensitive, likely because they can rely on coarse semantic cues rather than the precise ordering and alignment required by ranking-based objectives. Taken together, these results establish that code-switching challenges vary significantly by task demands, motivating targeted optimization beyond simple scale.

\begin{table*}[!t]
\centering
\small
\setlength{\tabcolsep}{5pt}

\resizebox{\textwidth}{!}{%
\begin{tabular}{l l *{5}{S[table-format=2.2]}}
\toprule
\textbf{Model} &
\textbf{Settings} &
{\textbf{Touch\'e 2020 (Argument)}} &
{\textbf{HumanEval (Code)}} &
{\textbf{TRECCOVID (Science)}} &
{\textbf{FollowIR (IF)}} &
{\textbf{Avg}} \\
\midrule

\multirow{4}{*}{\textit{all-MiniLM-L12-v2}} &
orig model + CSR-L-Chinese &
{23.85} & {60.37} & {39.51} & {-3.36} & {30.09} \\
& adapted model + CSR-L-Chinese &
{40.01} & {64.44} & {48.32} & {-1.87} & \textbf{37.73} \\
\cdashline{2-7}
& orig model + CSR-L-Japanese &
{22.12} & {62.18} & {36.12} & {-0.65} & {29.94} \\
& adapted model + CSR-L-Japanese &
{30.86} & {65.72} & {41.64} & {-0.88} & \textbf{34.34} \\
\midrule

\multirow{4}{*}{\textit{e5-large-v2}} &
orig model + CSR-L-Chinese &
{22.88} & {72.93} & {50.42} & {-4.97} & {35.32} \\
& adapted model + CSR-L-Chinese &
{38.55} & {74.18} & {64.26} & {-2.99} & \textbf{43.50} \\
\cdashline{2-7}
& orig model + CSR-L-Japanese &
{22.88} & {72.49} & {45.34} & {-1.93} & {34.70} \\
& adapted model + CSR-L-Japanese &
{26.98} & {76.77} & {56.96} & {-1.52} & \textbf{39.80} \\

\bottomrule
\end{tabular}%
}
\vspace{2pt}
\caption{Performance comparison of original and vocabulary-adapted models on the CSR-L benchmarks.}
\label{tab:vocab_expansion_results}
\end{table*}

\section{Vocabulary Expansion for Retrieval}
The CSR-L and CS-MTEB results establish two consistent observations: code-switching hurts end-task performance, and it also perturbs the representation space. This suggests that at least part of the failure arises at the input and encoding level. One plausible contributor is vocabulary and tokenization coverage: when tokens from a secondary language are split into many low-frequency subwords, the resulting representations can become noisy and drift away from the monolingual manifold. This motivates a controlled, low-cost intervention: lexicon-based vocabulary expansion, which extends the tokenizer with high-frequency missing words from the target language while leaving the main body of the retriever unchanged. If this intervention recovers a meaningful portion of the performance drop, it would indicate that vocabulary coverage is an important bottleneck; otherwise, it would imply that code-switching failures stem from deeper representation and training mismatches beyond the tokenizer. Within multilingual NLP, a significant body of work focuses on extending monolingual models to multilingual settings. For example, \citet{wang-etal-2022-expanding} adapted monolingual models to cover thousands of languages using lexicon-based techniques, and \citet{10.24963/ijcai.2023/698} introduced a vocabulary expansion algorithm designed to elicit robust multilingual performance from monolingual backbones. In this section, we ask whether the same idea can improve robustness to code-switched queries.

\subsection{Lexicon-Based Vocabulary Expansion}
To investigate whether bridging the semantic gap between languages can mitigate the performance degradation observed in code-switching retrieval, we implement a lexicon-based vocabulary expansion strategy. This method, adapted from the initialization techniques proposed by \citet{10.24963/ijcai.2023/698}, allows us to project the semantic capabilities of a well-aligned source language (e.g., English) onto the target language without requiring extensive multilingual pre-training.

Formally, we utilize an independent bilingual lexicon $\mathcal{D} = \{(w_t, w_s)\}$, which consists of word-level translation pairs mapping the target language to the source language. Distinct from this linguistic resource, the pre-trained source model operates on a subword vocabulary $\mathcal{V}_{pre}$ (e.g., WordPiece or BPE tokens) with a corresponding embedding matrix $\mathbf{E}_{pre} \in \mathbb{R}^{|\mathcal{V}_{pre}| \times d}$. Our objective is to initialize the embeddings for the target tokens $\mathcal{V}_{target}$ based on their semantic equivalents in the lexicon.

A significant challenge lies in the granularity mismatch: entries in the lexicon $\mathcal{D}$ are typically whole words, whereas the model vocabulary $\mathcal{V}_{pre}$ consists of subword units. To address this, we define the tokenizer as $T(\cdot)$ that decomposes a linguistic word $w$ into a sequence of subword tokens $[k_1, k_2, \dots, k_m]$, where each $k_i \in \mathcal{V}_{pre}$.

For a given target word $w_t$, we first identify its set of source translations $\mathcal{N}(w_t) = \{w_s \mid (w_t, w_s) \in \mathcal{D}\}$. To obtain a vector representation for a specific source word $w_s$, we tokenize it into its constituent subwords and average their pre-trained embeddings:

\begin{equation}
    \mathbf{v}_{w_s} = \frac{1}{|T(w_s)|} \sum_{k \in T(w_s)} \mathbf{e}_{k}
\end{equation}

where $\mathbf{e}_{k} \in \mathbf{E}_{pre}$ is the embedding of the subword token $k$. Finally, to initialize the embedding for the target token $w_t$, we aggregate the representations of all its valid source translations:

\begin{equation}
    \mathbf{e}_{w_t} = \frac{1}{|\mathcal{N}(w_t)|} \sum_{w_s \in \mathcal{N}(w_t)} \mathbf{v}_{w_s}
\end{equation}

In cases where a target token has no translation in the lexicon (i.e., $\mathcal{N}(w_t) = \emptyset$), we initialize $\mathbf{e}_{w_t}$ using a standard normal distribution $\mathcal{N}(0, \sigma^2)$. This hierarchical aggregation, from subword to word, then from source word to target word, ensures that the messy, fragmented nature of the pre-trained vocabulary does not hinder the effective transfer of semantic information to the code-switching context.

\subsection{Experiments and Results}
\paragraph{Experimental Setup} We apply our vocabulary expansion strategy to two representative English-only retrievers: \textit{all-MiniLM-L12-v2} and \textit{e5-large-v2}. To construct the semantic mapping, we utilize the high-quality bilingual lexicons provided by \citet{conneau2018wordtranslationparalleldata}. We independently expand these models to support both Chinese and Japanese, subsequently evaluating the performance of the adapted versions on our CSR-L benchmark.

\paragraph{Results} As shown in \autoref{tab:vocab_expansion_results}, lexicon-based vocabulary expansion consistently improves robustness to code-switching for both evaluated English-only retrievers. For \texttt{all-MiniLM-L12-v2}, adaptation increases the macro-average from 30.09 to 37.73 on CSR-L-Chinese and from 29.94 to 34.34 on CSR-L-Japanese, indicating a clear but partial recovery. Similarly, \texttt{e5-large-v2} benefits from adaptation, with the average improving from 35.32 to 43.50 (Chinese) and from 34.70 to 39.80 (Japanese). The gains are driven primarily by the two general retrieval benchmarks (e.g., Touch\'e 2020 and TRECCOVID), while improvements on HumanEval are comparatively smaller. Overall, these results mirror the earlier finding that code-switching imposes a substantial bottleneck, and demonstrate that vocabulary expansion provides a low-cost mitigation, but does not fully eliminate the deficit.

\section{Discussion}
In this paper, we identified code-switching as a persistent and universal bottleneck for modern IR. Across our constructed CSR-L and CS-MTEB benchmarks, we observed significant performance degradation regardless of whether the system employs statistical, dense, or late-interaction architectures. While multilingual training offers a degree of geometric stability, mitigating the severity of these drops compared to English-centric baselines, it fails to fully immunize models against the semantic disruption caused by mixed-language queries. Furthermore, our experiments with lexicon-based vocabulary expansion provide a nuanced insight: although this low-cost intervention yields measurable performance improvements, the resulting models still trail significantly behind English-only settings. This persistent gap underscores that code-switching is not merely a vocabulary coverage issue resolvable by surface-level patches, but a complex semantic challenge that necessitates dedicated architectural or training innovations to achieve true parity with monolingual systems.

\paragraph{Semantic Alignment vs. Retrieval Relevance} Our findings reveal a critical, often overlooked distinction between \textit{semantic alignment} and \textit{retrieval relevance} in code-switching contexts. While recent benchmarks such as \textbf{MINERS} \cite{winata-etal-2024-miners} demonstrate that multilingual models can achieve competitive performance in semantic tasks like bitext mining without fine-tuning, our results on CSR-L and CS-MTEB paint a significantly more complex picture. We observe that while current models effectively align synonyms across languages, which explains their resilience in simpler retrieval tasks, they struggle profoundly when tasked with the nuanced relevance modeling required for high-precision IR. This fragility is even preserved in large-scale foundation models; for instance, on CS-MTEB reranking tasks, the performance of \textit{Qwen3-Embedding-0.6B} plummets from a monolingual baseline of 63.09 to 37.33 in the Japanese setting. To sum up, these results underscore the immense heterogeneity across text embedding applications: proficiency in cross-lingual alignment does not guarantee robustness in code-switching IR, further validating the need for the specialized development of code-switching retrieval systems.

\paragraph{The Limits of Direct Multilingual Interaction}
\citet{zuo-etal-2025-evaluating} recently established that while LLMs excel as rerankers when inputs are translated (noisy monolingual IR), they fall severely short when interacting directly with multilingual bi-encoder outputs without intermediate translation. Our work extends this conclusion to the code-switching domain: just as models struggle with direct cross-lingual retrieval, they are similarly fragile when processing fluidly mixed-language queries. The failure of our vocabulary expansion experiments further corroborates this, indicating that surface-level fixes cannot compensate for the model's fundamental inability to process "native" mixed-language sequences. Collectively, these findings imply that future progress depends not on better translation or alignment but on developing training data that treat code-switching as a distinct linguistic modality.

\section*{Limitations}
We identify two main limitations in our work.

\noindent Language and phenomenon coverage. Our benchmarks operationalize code-switching as natural, query-level mixing between English and a small set of partner languages, which keeps the evaluation controlled and aligns with common search behavior where English technical terms appear inside otherwise non-English queries. At the same time, code-switching in the wild spans a broader space (e.g., romanization, transliteration and spelling variation, community-specific conventions, and mixed-language documents), which is not the primary focus of this study and remains a straightforward direction for future benchmark extensions.

\noindent Annotation and generation noise.
Working with code-switched text inevitably involves human judgment about what constitutes a natural switch while preserving the original information needs. We mitigate this through bilingual annotators, validation checks, conservative rewrite guidelines, and a manual spot check of generated CS-MTEB queries, but modest stylistic variation and occasional generation artifacts are difficult to eliminate entirely at scale. Accordingly, we emphasize consistent trends across models and settings, and release our resources to facilitate replication and expansion under alternative annotation or generation protocols.

\section*{Acknowledgments}
This work was supported by JST CRONOS (Grant Number JPMJCS25K5) and JST NEXUS (Grant Number JPMJNX25CA). Weihao Xuan is supported by RIKEN Junior Research Associate (JRA) Program.

\bibliography{custom}

\newpage
\appendix

\section{Instructions Given to Annotators for Rewriting the Queries}
\label{sec:instruction_to_annotator}

For Chinese CSR-L query rewriting, the annotators are native Chinese speakers with years of experience in English-speaking environments. They are asked to rewrite each original query into a natural code-switched form that reflects realistic conversational and search behavior. For Japanese CSR-L, the annotators are familiar with both English and Japanese usage, and the same instructions are applied. We reproduce the Chinese instructions below:
\begin{tcolorbox}[title=Instructions to Annotators]
\small
\textbf{Task:} Rewrite the given IR benchmark query into a \textbf{natural code-switched query} that mixes \textbf{English and the target language}, while keeping the \textbf{same information need}.
\\\\
\textbf{Instructions:}\\
- Read the original query and identify the exact information need (what the user wants to find).\\
- Rewrite it as a single query that a bilingual person might realistically type, mixing English and the target language.\\
- \textbf{Do not change meaning:} do not add new details or constraints; do not remove important details (time, location, stance, entities, domain terms).\\
- \textbf{Keep key terms stable:} keep named entities, product names, dataset identifiers, and technical terms unchanged unless there is a widely used and unambiguous target-language form.\\
- \textbf{Both languages must contribute:} avoid a fully monolingual query and avoid adding only one borrowed word. Each language should contain meaningful content.\\
- \textbf{Naturalness:} prefer phrase-level mixing (often keep technical keywords in English, express framing in the target language). Avoid word-by-word literal translation.\\
- \textbf{Length:} keep the rewrite roughly similar in length (about $\pm 20\%$). Do not add explanations or extra sentences.\\
- \textbf{If the query is about code:} do not translate code tokens, function names, variable names, operators, or error messages. Only rewrite the surrounding natural language.\\
- If you feel the query is impossible to rewrite to be natural code-switching queries, write \textbf{None}.\\
\\
\textbf{Example} \\
\begin{CJK*}{UTF8}{gbsn}Rewritten query: What are the causes and treatments of 神经性厌食症 and 神经性贪食症?\end{CJK*}\\
\\
\textbf{Now, here is the query awaiting rewritten:} \\
\{query\}
\end{tcolorbox}

\section{Additional Details on Evaluation}
\label{sec:evaluation_details}

Unless otherwise specified, all benchmark settings follow the standard MTEB evaluation process. Due to VRAM limitations, however, we set the batch size to 4 rather than 32 for all retrieval tasks except HumanEvalRetrieval. To speed up inference, we also enable FlashAttention 2 whenever the model supports it, which may lead to slight differences from the public MTEB leaderboard. The metrics used for each task type are as follows:
\begin{itemize}
    \item Retrieval: nDCG@10
    \item Instruction Reranking: $p$-MRR \cite{weller-etal-2025-followir}
    \item Clustering: V-measure \cite{rosenberg-hirschberg-2007-v}
    \item Classification: Accuracy
    \item STS: Cosine Spearman correlation
    \item Reranking: MAP@1000
    \item Pair Classification: mean average precision
\end{itemize}

\section{CSR-L Results on Japanese}
\label{sec:csrl_results_on_japanese}
The results for CSR-L-Japanese are presented in \autoref{tab:csrl_results_ndcg10_avg_japanese}.
\begin{table*}[!t]
\centering
\small
\setlength{\tabcolsep}{5pt}

\resizebox{\textwidth}{!}{%
\begin{tabular}{l l *{8}{S[table-format=2.2]} *{2}{S[table-format=2.2]} S[table-format=-2.2]}
\toprule
\multirow{2}{*}{\textbf{Method Family}} &
\multirow{2}{*}{\textbf{Model}} &
\multicolumn{2}{c}{\textbf{Touch\'e 2020}} &
\multicolumn{2}{c}{\textbf{HumanEval}} &
\multicolumn{2}{c}{\textbf{TRECCOVID}} &
\multicolumn{2}{c}{\textbf{FollowIR}} &
\multicolumn{1}{c}{\textbf{Avg}} &
\multicolumn{1}{c}{\textbf{Avg}} &
\multicolumn{1}{c}{\textbf{Drop}} \\
\cmidrule(lr){3-4}\cmidrule(lr){5-6}\cmidrule(lr){7-8}\cmidrule(lr){9-10}\cmidrule(lr){11-13}
& &
{\textbf{Orig}} & {\textbf{CSR-L}} &
{\textbf{Orig}} & {\textbf{CSR-L}} &
{\textbf{Orig}} & {\textbf{CSR-L}} &
{\textbf{Orig}} & {\textbf{CSR-L}} &
{\textbf{Orig}} & {\textbf{CSR-L}} &
{$\Delta$} \\
\midrule

\multirow{1}{*}{Statistical} &
BM25 &
{60.32} & {39.17} & {35.02} & {34.75} & {55.62} & {43.48} & {-0.62} & {0.48} & {37.59} & {29.47} & {-8.12} \\
\midrule

\multirow{9}{*}{Bi-encoder} &
\textit{e5-large-v2} &
{42.52} & {22.88} & {80.70} & {72.49} & {66.64} & {45.34} & {-0.99} & {-1.93} & {47.22} & {34.70} & {-12.52} \\
& \textit{all-MiniLM-L12-v2} &
{49.22} & {22.12} & {70.08} & {62.18} & {51.17} & {36.12} & {-0.66} & {-0.65} & {42.45} & {29.94} & {-12.51} \\
& \textit{mE5-large} &
{49.32} & {40.54} & {81.15} & {75.28} & {71.56} & {53.85} & {-3.38} & {-2.26} & {49.66} & {41.85} & {-7.81} \\
& \textit{bge-m3} &
{55.02} & {45.53} & {61.33} & {58.85} & {54.70} & {45.41} & {-2.94} & {-2.84} & {42.03} & {36.74} & {-5.29} \\
& \textit{Arctic-Embed-m-v2.0} &
{65.29} & {47.89} & {78.70} & {74.29} & {80.45} & {73.35} & {-3.20} & {-3.13} & {55.31} & {48.10} & {-7.21} \\
& \textit{Arctic-Embed-l-v2.0} &
{64.05} & {53.19} & {71.27} & {70.60} & {83.63} & {79.07} & {-2.45} & {-2.04} & {54.13} & {50.21} & {-3.92} \\
& \textit{Qwen3-Embedding-0.6B} &
{71.65} & {58.77} & {94.24} & {94.80} & {89.43} & {78.63} & {5.10} & {4.11} & {65.11} & {59.08} & {-6.03} \\
& \textit{Qwen3-Embedding-4B} &
{75.07} & {60.33} & {98.12} & {96.15} & {92.95} & {88.11} & {11.87} & {11.26} & {69.50} & {63.96} & {-5.54} \\
& \textit{Qwen3-Embedding-8B} &
{75.77} & {68.69} & {99.22} & {98.52} & {94.68} & {88.14} & {9.86} & {8.73} & {69.88} & {66.02} & {-3.86} \\
\midrule

\multirow{3}{*}{Cross-encoder} &
\textit{jina-reranker-v3} &
{22.68} & {24.49} & {85.53} & {83.41} & {81.32} & {71.28} & {-0.27} & {-0.03} & {47.32} & {44.79} & {-2.53} \\
& \textit{bge-reranker-v2-m3} &
{35.48} & {28.94} & {43.74} & {39.91} & {79.00} & {66.14} & {-1.38} & {-0.76} & {39.21} & {33.56} & {-5.65} \\
& \textit{Qwen3-Reranker-0.6B} &
{29.15} & {23.93} & {83.74} & {81.90} & {84.30} & {72.80} & {1.40} & {0.55} & {49.65} & {44.80} & {-4.85} \\
& \textit{Qwen3-Reranker-4B} & {37.76} & {27.96} & {85.29} & {83.75} & {85.44} & {73.81} & {2.33} & {0.55} & {52.71} & {46.52} & {-6.19} \\
& \textit{Qwen3-Reranker-8B} & {40.91} & {32.21} & {85.53} & {84.26} & {84.58} & {71.71} & {2.74} & {1.22} & {53.44} & {47.35} & {-6.09} \\
\midrule

\multirow{1}{*}{Late-interaction} &
ColBERT v2 &
{61.62} & {31.18} & {40.30} & {34.23} & {69.30} & {48.86} & {-0.95} & {0.87} & {42.57} & {28.79} & {-13.78} \\

\bottomrule
\end{tabular}%
}

\vspace{2pt}
\caption{nDCG@10 and \textit{p-MRR} on the original (Orig) and code-switched (CSR-L) queries across four IR benchmarks on English-Japanese code-switching. Avg is the macro-average over the four datasets. Drop $\Delta$ is computed as Avg(CSR-L) - Avg(Orig); negative values indicate performance degradation under code-switching.}
\label{tab:csrl_results_ndcg10_avg_japanese}
\end{table*}

\section{Prompt Template for Doing Code-switching}
\label{appendix:prompt_template_doing_code_switching}

We take the prompt for CSR-L-Chinese tasks as an example, shown below:

\begin{tcolorbox}[breakable, title=Prompt Template: English-Chinese Code-Switching Rewrite]
\small

\textbf{Role:} You rewrite English paragraphs/sentences into natural English--Chinese code-switched texts, like something a bilingual Mainland Chinese/English user would type. If the input is a paragraph, you should do \textbf{\{num\_cs\}} code-switching changes in the original text.

\textbf{Task:}
\begin{itemize}
  \item \textbf{Input:} one dense English paragraph/sentence (for search).
  \item \textbf{Output:} one single English-Chinese mixed paragraph/sentence.
\end{itemize}

\textbf{Style \& rules:}

\begin{enumerate}
  \item \textbf{Base language}
  \begin{itemize}
    \item The query should be mainly in English.
    \item Use Simplified Chinese only where Chinese is the more natural or widely used expression for Chinese users.
  \end{itemize}

  \item \textbf{Preserve meaning}
  \begin{itemize}
    \item Keep all important keywords and intent from the original sentence.
    \item Keep product names, library names, frameworks, and proper nouns in English (e.g., Python, React, Kubernetes, Notion, Apple, RTX 4060), unless there is an extremely standard Chinese name everyone uses.
  \end{itemize}

  \item \textbf{When to use Chinese} (use Chinese especially for idioms, for example)
  \begin{itemize}
    \item Learning/guide phrases:
      \zh{入门教程}, \zh{入门课程}, \zh{学习路线}, \zh{使用教程}, \zh{实战教程}, \zh{完整指南}, \zh{保姆级教程},
      \zh{选购指南}, \zh{选购建议}, \zh{配置推荐}, \zh{排行榜}, \zh{使用心得}, \zh{避坑}, \zh{踩坑经验}, \zh{速查表}.
    \item Evaluation/preference words:
      \zh{性价比}, \zh{便宜一点}, \zh{高性价比}, \zh{对比}, \zh{推荐}, \zh{怎么选}, \zh{适合新手}, \zh{适合大学生}.
    \item Very common Chinese net-terms that replace simple English nouns, for example:
      \begin{itemize}
        \item ``gaming laptop'' $\rightarrow$ \zh{游戏本}
        \item Keep most other general nouns in English unless the Chinese term is clearly more common among Chinese users.
      \end{itemize}
  \end{itemize}

  \item \textbf{How to mix}
  \begin{itemize}
    \item Keep the overall structure and most content words in English.
    \item Insert Chinese at the phrase level (e.g., ``... \zh{入门教程}'', ``... \zh{性价比对比}'', ``best \zh{游戏本} for college students \zh{性价比}''), not every other word.
    \item Keep it short: do not turn it into a full explanation.
  \end{itemize}

  \item \textbf{Do NOT}
  \begin{itemize}
    \item Do NOT translate the whole sentence into Chinese.
    \item Do NOT restate the sentence twice in different languages.
    \item Do NOT add romanization or language labels (no ``[in Chinese]'', etc.).
    \item Do NOT add commentary or explanation.
    \item Do NOT use emoji to do code switching.
    \item Do NOT output any chain-of-thought, reasoning, or thinking process.
    \item Do NOT include any preamble like ``Here is the rewritten version:''.
  \end{itemize}

  \item \textbf{Be faithful}
  \begin{itemize}
    \item Keep the same length as the original paragraph/sentence.
    \item Keep the same sentence order as the original input.
  \end{itemize}

  \item \textbf{Output format}
  \begin{itemize}
    \item Your response MUST start directly with \texttt{<code\_switched\_output>} and end with \texttt{</code\_switched\_output>}.
    \item Output ONLY the XML tags with your answer inside. Nothing else.
    \item No text before or after the XML tags.
  \end{itemize}
\end{enumerate}

\textbf{Output XML format:}\\
\texttt{<code\_switched\_output>Your answer</code\_switched\_output>}

\vspace{2pt}
\textbf{Examples of desired style (do NOT output these literally):}
\begin{itemize}
  \item \texttt{Python data analysis } \zh{入门教程}
  \item \texttt{best } \zh{游戏本} \texttt{ for college students } \zh{性价比}
  \item \texttt{Notion personal knowledge management } \zh{入门教程}
  \item \texttt{markdown } \zh{速查表}
\end{itemize}

\textbf{Keep the similar length (measured by number of sentences) as the original input. Do not over-simplify the query.}

\vspace{2pt}
\textbf{Now rewrite this paragraph/sentence into a natural English--Chinese code-switched version:}\\
\{your\_query\_here\}

\end{tcolorbox}

\section{CSR-L-Chinese Query Examples}

Examples of rewritten code-switching queries in CSR-L-Chinese are listed in tables below from \autoref{tab:touche2020_cs_example_cn} to \autoref{tab:followIR_cs_example_cn}.

\begin{table}[!t]
\centering
\begin{tabular}{@{} p{\columnwidth} @{}}
\toprule
\textit{Query} \\
\midrule
Should teachers get \zh{终身教职}?\\
\bottomrule
\caption{CSR-L-Chinese Touché 2020 Code-Switching Example.}
\label{tab:touche2020_cs_example_cn}
\end{tabular}
\end{table}

\begin{table}[!t]
\centering
\begin{tabular}{@{} p{\columnwidth} @{}}
\toprule
\textit{Query} \\
\midrule
Given an array of non-negative integers, return a sorted copy: if sum(first index value, last index value) is odd sort ascending, if even sort descending. \zh{注意: 不要修改原数组。示例}: sort\_array([]) => [], sort\_array([5]) => [5], sort\_array([2,4,3,0,1,5]) => [0,1,2,3,4,5]?\\
\bottomrule
\caption{CSR-L-Chinese HumanEval Code-Switching Query Example.}
\label{tab:humanEval_cs_example_cn}
\end{tabular}
\end{table}

\begin{table}[!t]
\centering
\begin{tabular}{@{} p{\columnwidth} @{}}
\toprule
\textit{Query} \\
\midrule
Will SARS-CoV2 infected people develop immunity\zh{，交叉保护是否可能？}\\
\bottomrule
\caption{CSR-L-Chinese TRECCOVID Code-Switching Query Example.}
\label{tab:treccovid_cs_example_cn}
\end{tabular}
\end{table}

\begin{table}[!t]
\centering
\begin{tabular}{@{} p{\columnwidth} @{}}
\toprule
\textit{Query-og} \\
\midrule
What efforts have been made to stabilize the \zh{比萨斜塔}, and how successful have the efforts been?\\
\bottomrule
\textit{Instruction-og} \\
\midrule
Relevant documents provide discussions of the current condition of the tower, describe the \zh{加固措施} taken, and/or provide measurements reflecting change in the tower.\\
\bottomrule
\textit{Query-changed} \\
\midrule
What efforts have been made to stabilize the \zh{比萨斜塔}, and how successful have the efforts been?\\
\bottomrule
\textit{Instruction-changed} \\
\midrule
Relevant documents provide discussions of the current condition of the tower, describe the \zh{加固措施} taken, and/or provide measurements reflecting change in the tower. Exclude documents mentioning the year 1990.\\
\bottomrule
\caption{CSR-L-Chinese FollowIR Code-Switching Query And Instruction Example. It conforms to the original format of MTEB, which uses the same query as query-og and query-changed but has difference in instruction-og and instruction-changed.}
\label{tab:followIR_cs_example_cn}
\end{tabular}
\end{table}

\section{Japanese-CSR-L Statistics And Query Examples}

Statistics and the examples of rewritten code-switching queries in Japanese-CSR-L are listed in tables below from \autoref{tab:statistics_of_Japanese_CSRL} to \autoref{tab:followIR_cs_example_jp}.

\begin{table}[!t]
\centering
\resizebox{\columnwidth}{!}{
\begin{tabular}{l|rrr|rr|c}
\toprule
& \multicolumn{3}{c|}{\textbf{Total Number}}
& \multicolumn{2}{c|}{\textbf{Avg. Length}}
& \textbf{Examples} \\
\cmidrule{2-4}\cmidrule{5-6}\cmidrule{7-7}
\textbf{Dataset}
& $\mathbf{Q}$
& $\boldsymbol{\mathcal{D}}$
& $\boldsymbol{\mathcal{D}^{+}}$
& $\mathbf{Q}$
& $\boldsymbol{\mathcal{D}}$
&  \\
\midrule
Touch\'e 2020     & 49 & 303,732 & 34.94 & 16.39 & 451.51 & \autoref{tab:touche2020_cs_example_jp} \\
HumanEval       & 158 & 158 & 1.00 & 88.48 & 98.20 & \autoref{tab:humanEval_cs_example_jp} \\
TRECCOVID   & 50 & 171,332 & 493.46 & 22.98 & 223.51 & \autoref{tab:treccovid_cs_example_jp} \\
FollowIR  & 208 & 98,312 & 30.00 & 120.86 & 465.39 & \autoref{tab:followIR_cs_example_jp} \\
\bottomrule
\end{tabular}
}
\caption{Statistics of datasets in Japanese-CSR-L. \textit{Q}: number of queries; \textit{D}: corpus size; \textit{D}$^+$: average positive documents per query. Avg. Length is measured in tokens. Examples can be seen in the tables in Appendix.}
\label{tab:statistics_of_Japanese_CSRL}
\end{table}

\begin{table}[!t]
\centering
\begin{tabular}{@{} p{\columnwidth} @{}}
\toprule
\textit{Query} \\
\midrule
Do violent video games contribute to \ja{若者の暴力}?\\
\bottomrule
\caption{Japanese-CSR-L Touché 2020 Code-Switching Example.}
\label{tab:touche2020_cs_example_jp}
\end{tabular}
\end{table}

\begin{table}[!t]
\centering
\begin{tabular}{@{} p{\columnwidth} @{}}
\toprule
\textit{Query} \\
\midrule
Given a string, find out how many distinct characters (\ja{大文字・小文字を問わず}) does it consist of\\
\bottomrule
\caption{Japanese-CSR-L HumanEval Code-Switching Query Example.}
\label{tab:humanEval_cs_example_jp}
\end{tabular}
\end{table}

\begin{table}[!t]
\centering
\begin{tabular}{@{} p{\columnwidth} @{}}
\toprule
\textit{Query} \\
\midrule
best masks for Covid-19 \ja{感染予防 おすすめ}\\
\bottomrule
\caption{Japanese-CSR-L TRECCOVID Code-Switching Query Example.}
\label{tab:treccovid_cs_example_jp}
\end{tabular}
\end{table}

\begin{table}[!t]
\centering
\begin{tabular}{@{} p{\columnwidth} @{}}
\toprule
\textit{Query-og} \\
\midrule
What standards do cruise ships use for \ja{衛生と安全の維持}?\\
\bottomrule
\textit{Instruction-og} \\
\midrule
Relevant documents refer to \ja{衛生と安全} practices and standards for \ja{レジャークルーズ船}. Not relevant are standards for small pleasure craft or commercial freight ships, tankers, etc. Documents referring to a specific ship's problems are not relevant.\\
\bottomrule
\textit{Query-changed} \\
\midrule
What standards do cruise ships use for \ja{衛生と安全の維持}?\\
\bottomrule
\textit{Instruction-changed} \\
\midrule
Relevant documents refer to \ja{衛生と安全} practices and standards for \ja{レジャークルーズ船}, but don't include information about Royal Caribbean or Royal Viking. Not relevant are standards for small pleasure craft or commercial freight ships, tankers, etc. Documents referring to a specific ship's problems are not relevant.\\
\bottomrule
\caption{Japanese-CSR-L FollowIR Code-Switching Query And Instruction Example. It conforms to the original format of MTEB, which uses the same query as query-og and query-changed but has difference in instruction-og and instruction-changed.}
\label{tab:followIR_cs_example_jp}
\end{tabular}
\end{table}

\section{Additional Results on CS-MTEB}
In \autoref{tab:cs-mteb-additional-five-languages}, we report CS-MTEB results on additional five languages.

\begin{table*}[t]
\centering
\small
\setlength{\tabcolsep}{4pt}
\resizebox{\textwidth}{!}{
\begin{tabular}{llcccccccc}
\toprule
\textbf{Model} & \textbf{Setting} &
\textbf{Instr. Rerank} (1) &
\textbf{Retrieval} (5) &
\textbf{Clust.} (1) &
\textbf{Cls.} (1) &
\textbf{STS} (1) &
\textbf{Rerank} (1) &
\textbf{Pair Cls.} (1) &
\textbf{Total} (11) \\
\midrule

\multirow{6}{*}{e5-large-v2}
& Original & -0.99 & 51.78 & 62.00 & 73.97 & 84.55 & 60.17 & 59.88 & 55.91 \\
\cdashline{2-10}
& Korean    & -0.50	& 37.92	& 24.78	& 64.43	& 55.52	& 61.25	& 55.01	& 42.63 \\
& French    & -0.8  & 44.18 & 26.29	& 70.28	& 57.6  & 64.97	& 52.57	& 45.01 \\
& Italian   & -0.68	& 40.93	& 24.72	& 65.14	& 56.1	& 60.32	& 51.63	& 42.59 \\
& Portuguese& -1.61 & 43.39 & 25.68	& 67.83	& 56.37	& 62.72	& 56.85	& 44.46 \\
& Dutch     & -2.11	& 39.93 & 26.28	& 64.26	& 55.1	& 62.02	& 58.2	& 43.38 \\
\midrule

\multirow{6}{*}{Arctic-Embed-m-v2.0}
& Original & -3.20 & 64.21 & 60.09 & 64.78 & 75.97 & 62.37 & 58.09 & 54.62 \\
\cdashline{2-10}
& Korean    & -2.52	& 55.26	& 36.7  & 61.29 & 55.89 & 59.21 & 53.57 & 45.63  \\
& French    & -3.18	& 60.85 & 36.53 & 66.61	& 57.63	& 62.71	& 52.47	& 47.66  \\
& Italian   & -3.53	& 61.50 & 37.64	& 66.85	& 57.34	& 62.41	& 50.21	& 47.49  \\
& Portuguese& -2.94 & 61.29	& 37.19	& 66.47	& 56.12	& 61.89	& 58.12	& 48.31 \\
& Dutch     & -3.97	& 58.13	& 36.63	& 65.14	& 54.99	& 60.02	& 55.89	& 46.69  \\
\midrule

\multirow{6}{*}{Qwen3-Embedding-0.6B}
& Original & 5.10 & 73.67 & 68.21 & 72.07 & 91.14 & 63.09 & 75.55 & 64.12 \\
\cdashline{2-10}
& Korean    & 2.72 & 67.34	& 36.82	& 84.08	& 73.17	& 67.73	& 59.33	& 55.89 \\
& French    & 2.28 & 67.97	& 35.33	& 85.57	& 73.18	& 68.46	& 59.31	& 56.02  \\
& Italian   & 3.34 & 68.85 	& 35.49	& 83.74	& 72.29	& 68.6	& 58.64	& 55.85 \\
& Portuguese& 4.84 & 69.55	& 35.51	& 84.69	& 72.2	& 67.9	& 64.08	& 56.97  \\
& Dutch     & 2.42 & 65.87	& 36.25	& 82.91	& 70.83	& 66.32	& 63.54	& 55.45  \\
\bottomrule
\end{tabular}
}
\caption{CS-MTEB results by model and evaluation setting. Columns correspond to CS-MTEB task categories, with the number of tasks per category in parentheses. The result is the macro average over 7 task categories / Mean (TaskType).}
\label{tab:cs-mteb-additional-five-languages}
\end{table*}

\section{Additional Discussion on a Newly Curated Retrieval Benchmark}
To further probe whether the CSR-L findings are overly tied to well-known benchmark suites, we add an extra retrieval-only check on \textit{AILACaseDocs}, a dataset that was newly introduced in the recent RTEB leaderboard and is not part of the commonly used MMTEB leaderboard. Following the same query-side prompting procedure described for CS-MTEB, we construct a Chinese code-switched version and evaluate \textit{mE5-large} together with \textit{Qwen3-Embedding-0.6B} under the same protocol.

\begin{table}[!t]
\centering
\small
\begin{tabular}{l S[table-format=2.2] S[table-format=2.2] S[table-format=-2.2]}
\toprule
\textbf{Model} & {\textbf{Orig}} & {\textbf{Chinese}} & {\textbf{Drop}} \\
\midrule
\textit{mE5-large} & {41.89} & {23.83} & {-18.06} \\
\textit{Qwen3-Embedding-0.6B} & {34.80} & {31.85} & {-2.95} \\
\bottomrule
\end{tabular}
\caption{Additional results on AILACaseDocs. Drop is computed as Chinese - Orig.}
\label{tab:ailacasedocs_additional_results}
\end{table}

The results in \autoref{tab:ailacasedocs_additional_results} show that the same phenomenon persists on this newer benchmark: both models degrade when the query is code-switched, with the drop being substantial for \textit{mE5-large} and smaller but still non-trivial for \textit{Qwen3-Embedding-0.6B}. While we do not claim that \textit{AILACaseDocs} is completely isolated from broader benchmark-ecosystem effects, this additional check reduces the concern that our conclusions are driven solely by in-domain adaptation to a small set of long-standing public evaluation datasets.

\section{Additional Discussion on Two-Stage Reranking}
Because the CSR-L cross-encoder results in the main tables are obtained by direct full-corpus scoring, we additionally test a standard two-stage setup on CSR-L-Chinese. Specifically, we use \textit{Qwen3-Embedding-0.6B} as the first-stage retriever, keep the top-100 candidates for each query, and then rerank them with \textit{jina-reranker-v3} or \textit{Qwen3-Reranker-0.6B}.

\begin{table}[!t]
\centering
\small
\setlength{\tabcolsep}{3pt}
\resizebox{\columnwidth}{!}{
\begin{tabular}{l *{8}{S[table-format=2.2]} S[table-format=2.2] S[table-format=2.2] S[table-format=-2.2]}
\toprule
\textbf{Model} &
{\textbf{Touch\'e O}} & {\textbf{Touch\'e C}} &
{\textbf{HE O}} & {\textbf{HE C}} &
{\textbf{TREC O}} & {\textbf{TREC C}} &
{\textbf{FIR O}} & {\textbf{FIR C}} &
{\textbf{Avg O}} & {\textbf{Avg C}} & {\textbf{Drop}} \\
\midrule
\textit{jina-reranker-v3} & {62.04} & {58.21} & {98.22} & {98.07} & {89.37} & {84.20} & {4.65} & {3.13} & {63.57} & {60.90} & {-2.67} \\
\textit{Qwen3-Reranker-0.6B} & {73.05} & {67.65} & {97.33} & {97.29} & {91.69} & {88.36} & {0.08} & {0.93} & {65.54} & {63.56} & {-1.98} \\
\bottomrule
\end{tabular}
}
\caption{Additional two-stage reranking results on CSR-L-Chinese. O/C denote original and Chinese code-switched queries, respectively. Drop is computed as Avg C - Avg O.}
\label{tab:csrl_two_stage_reranking}
\end{table}

The results in \autoref{tab:csrl_two_stage_reranking} show that the same code-switching degradation persists under a strong and standard retrieval pipeline: even after retrieving with \textit{Qwen3-Embedding-0.6B} and reranking the top-100 candidates, both rerankers still perform worse on the code-switched queries than on the original English ones. This confirms that the performance drop is not an artifact of the direct full-corpus cross-encoder setup alone.

\section{Additional Discussion on Non-English Monolingual Baselines}
To separate code-switching effects from simply moving away from English, we add a Chinese-centric evaluation in which both the monolingual baseline queries and the document collection are Chinese. Concretely, we use the Chinese subset of \textit{MIRACLRetrievalHardNegatives}, then convert the original Chinese queries into Chinese--English code-switched queries with the same prompting procedure while keeping the documents unchanged.

\begin{table}[!t]
\centering
\small
\begin{tabular}{l S[table-format=2.2] S[table-format=2.2] S[table-format=-2.2]}
\toprule
\textbf{Model} & {\textbf{Orig}} & {\textbf{CS}} & {\textbf{Drop}} \\
\midrule
\textit{jina-embeddings-v3} & {57.89} & {50.50} & {-7.39} \\
\textit{Qwen3-Embedding-0.6B} & {60.19} & {55.56} & {-4.63} \\
\textit{Arctic-Embed-l-v2.0} & {61.18} & {53.30} & {-7.88} \\
\bottomrule
\end{tabular}
\caption{Additional results on the Chinese subset of MIRACLRetrievalHardNegatives. CS denotes Chinese--English code-switched queries, and Drop is computed as CS - Orig.}
\label{tab:miracl_chinese_codeswitching}
\end{table}

The results in \autoref{tab:miracl_chinese_codeswitching} show that the degradation persists even when the monolingual baseline is non-English: all three retrievers perform worse on the code-switched queries than on the original Chinese ones. This indicates that the effect we observe is not merely a consequence of moving away from English as the highest-resource language, but also appears in a Chinese-centric retrieval setting where the comparison axis is monolingual Chinese versus Chinese--English code-switching.

\section{Additional Discussion on Query Quality Verification}
To provide a direct quality check for the automatically generated CS-MTEB queries, we manually inspected 50 sampled rewritten queries. Two raters independently scored each query on a 1--10 scale along two axes: \textit{naturalness}, which measures whether the code-switching pattern resembles a plausible bilingual user query, and \textit{information preservation}, which measures whether the rewritten query retains the original information need.

\begin{table}[!t]
\centering
\small
\begin{tabular}{l S[table-format=1.2] S[table-format=1.2] S[table-format=1.2]}
\toprule
\textbf{Criterion} & {\textbf{Rater 1}} & {\textbf{Rater 2}} & {\textbf{Mean}} \\
\midrule
\textit{Naturalness} & {9.02} & {9.30} & {9.16} \\
\textit{Information Preservation} & {9.80} & {9.76} & {9.78} \\
\bottomrule
\end{tabular}
\caption{Manual quality check on 50 sampled CS-MTEB rewritten queries. Each query is rated independently by two raters on a 1--10 scale.}
\label{tab:csmteb_query_quality_check}
\end{table}

As shown in \autoref{tab:csmteb_query_quality_check}, the sampled queries receive high scores from both raters on both dimensions. In particular, information preservation is consistently close to the ceiling, indicating that the rewritten queries largely maintain the original search intent, while the naturalness scores also remain high, suggesting that the inserted language switches are generally fluent and plausible. Although this spot check does not replace full-scale human verification, it provides additional evidence that the automatic rewriting procedure yields sufficiently reliable queries for benchmark construction.

\section{GenAI Statement}
This work utilized generative AI tools to assist with formatting, generating LaTeX templates, and refining word choice. The authors reviewed and verified all AI-assisted content to ensure factual accuracy and academic integrity.

\section{License Statement}
In this project, we use the MTEB evaluation framework \cite{muennighoff2023mtebmassivetextembedding}, which is released under the Apache License 2.0. Our evaluation datasets are largely accessed through the MTEB suite and their original sources (for example, the Hugging Face Hub); each dataset is used in accordance with its respective license terms.

We also use the following publicly released model checkpoints under their stated licenses: \textit{all-MiniLM-L12-v2} (Apache License 2.0), \textit{e5-large-v2} (MIT License), \textit{Arctic-Embed-m/l-v2.0} (Apache License 2.0), \textit{Qwen3-Embedding-0.6/4/8B} (Apache License 2.0), \textit{jina-reranker-v3} (CC BY-NC 4.0), \textit{bge-reranker-v2-m3} (Apache License 2.0), \textit{Qwen3-Reranker-0.6/4/8B} (Apache License 2.0), and \textit{ColBERT v2} (MIT License). We use these models for research and evaluation purposes and comply with the corresponding license requirements (including non-commercial restrictions where applicable).

\end{document}